\documentclass[prb,superscriptaddress,twocolumn]{revtex4}
\usepackage[utf8]{inputenc}
\usepackage{graphicx}
\usepackage{dcolumn}
\usepackage{bm}
\usepackage{color}
\usepackage{soul}
\usepackage{times}

\begin{document}

\title{Ambient temperature high-pressure-induced ferroelectric phase transition in CaMnTi$_2$O$_6$}

\author{J. Ruiz-Fuertes}
\email{javier.ruiz-fuertes@uv.es}
\affiliation{MALTA-Consolider Team, Departament de F\'{i}sica Aplicada-ICMUV, Universitat de Val\`{e}ncia, Dr. Moliner 50, 46100 Burjassot, Valencia, Spain}
\author{T. Bernert}
\affiliation{Max-Planck-Institut f\"ur Kohlenforschung, Kaiser-Wilhelm-Platz 1, D-45470 M\"ulheim an der Ruhr, Germany}
\author{D. Zimmer}
\author{N. Schrodt}
\affiliation{Institut f\"ur Geowissenschaften, Goethe-Universit\"at, Altenh\"oferallee 1, 60438 Frankfurt am Main, Germany}
\author{M. Koch-M\"uller}
\affiliation{GFZ Potsdam, Sektion 4.3, Telegrafenberg, 14473 Potsdam, Germany}
\author{B. Winkler}
\author{L. Bayarjargal}
\affiliation{Institut f\"ur Geowissenschaften, Goethe-Universit\"at, Altenh\"oferallee 1, 60438 Frankfurt am Main, Germany}
\author{C. Popescu}
\affiliation{CELLS-ALBA Synchrotron Light Facility, 08290 Cerdanyola del Vallés, Barcelona, Spain}
\author{S. MacLeod}
\affiliation{Atomic Weapons Establishment, Aldermaston, Reading, RG7 4PR, UK}
\affiliation{Institute of Shock Physics, Imperial College London, London, SW7 2AZ, UK}
\author{K. Glazyrin}
 \affiliation {Deutsches Elektronen-Synchrotron DESY, Notkestrasse 85, D-22603 Hamburg, Germany}
\date{\today}

\begin{abstract}
The ferroelectric to paraelectric phase transition of
multiferroic CaMnTi$_2$O$_6$ has been investigated at high pressures
and ambient temperature by second harmonic generation (SHG),
Raman spectroscopy, and powder and single-crystal x-ray
diffraction. We have found that CaMnTi$_2$O$_6$ undergoes  a
pressure-induced structural phase transition
($P4_2mc \rightarrow P4_2/nmc$) at $\sim$7 GPa to the
same paraelectric structure found at ambient pressure and $T_c$ = 630
K. The continuous linear decrease of the SHG intensity that disappears at 7 GPa and the existence of a Raman active mode at 244 cm$^{-1}$ that first softens
up to 7 GPa and then hardens with pressure, are used to discuss the nature of the phase transition of CaMnTi$_2$O$_6$ for which a d$T_c$/d$P = -48$ K/GPa has been found. Neither
a volume contraction nor a change of  the normalized pressure on the
eulerian strain are observed across the phase transition with all the
unit-cell volume data following a second order Birch-Murnaghan
equation of state with a bulk modulus of $B_0$ = 182.95(2) GPa.

\end{abstract}

\maketitle

\section{INTRODUCTION}
\label{int}

Multiferroics have attracted a significant amount of attention as they
potentially offer the prospect of doubling the capacity of information
storage in combined MRAM-FRAM memories \cite{cheon07}.
The largest coexisting ferroelectric and magnetic effects are found in
type I multiferroics, in which an ambient temperature ferroelectric
compound containing a magnetic ion shows an antiferromagnetic order at
sufficiently low temperature. Unfortunately, while most type I
multiferroics have a large spontaneous polarization, they do not show
bulk ferroelectricity due to their large leakage and coercive field
\citep{aimia11}. Recently, \citet{aimia14} have discovered that
CaMnTi$_2$O$_6$ is a multiferroic with a moderate spontaneous
polarization of $\sim$24 $\mu$C/cm$^2$ and the first example of an
oxide containing Mn$^{2+}$ allowing a polarization reversal at ambient
temperature, thus leading to a novel class of type I multiferroics.

\begin{figure}
\centering
\includegraphics[width=0.3\textwidth]{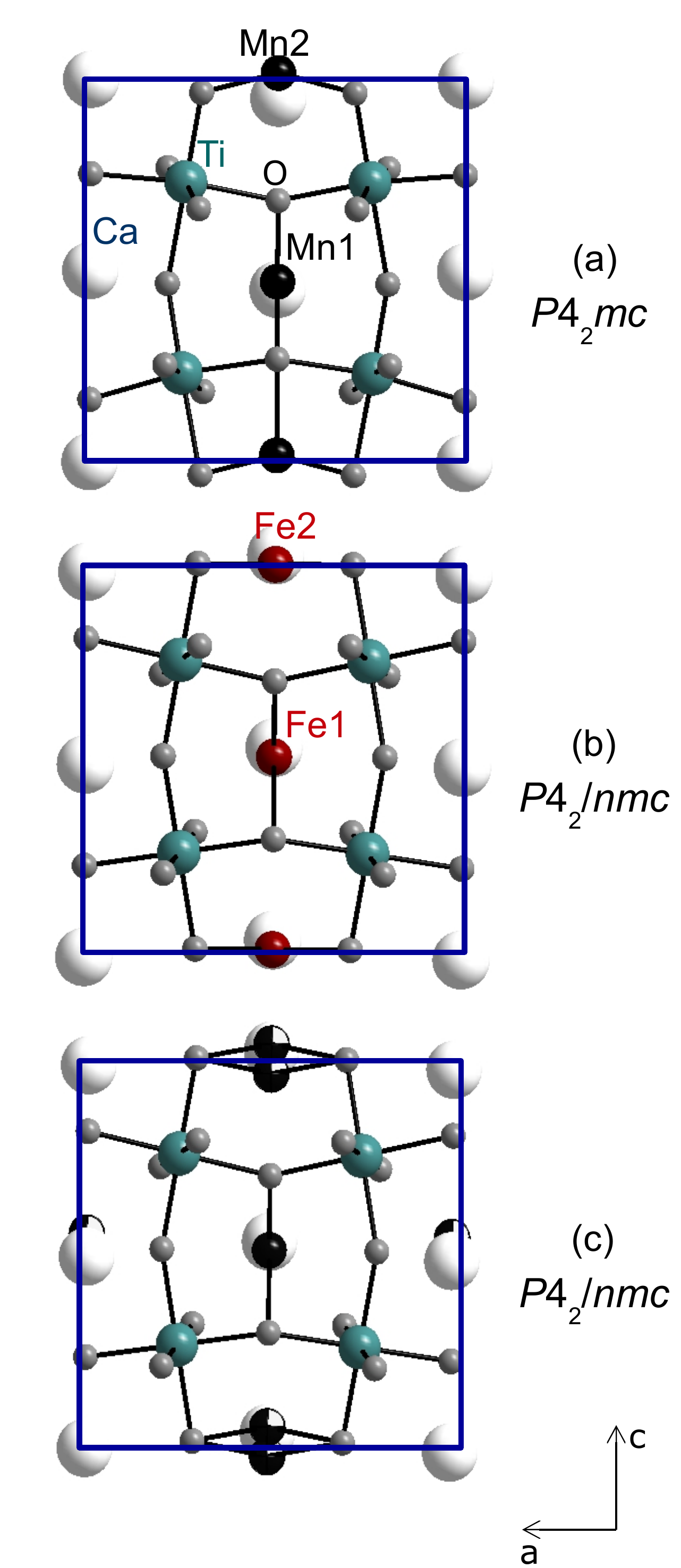}
\caption{\label{fig:fig1} Projection of the A-site ordered perovskite
  structure projected along the [010] direction for (a) the
  ferroelectric CaMnTi$_2$O$_6$ (S. G. $P4_2mc$), (b) CaFeTi$_2$O$_6$ (S. G. $P4_2/nmc$), and (c)
  paraelectric CaMnTi$_2$O$_6$ (S. G. $P4_2/nmc$).}
\end{figure}

CaMnTi$_2$O$_6$ crystallizes in a tetragonal A-site-ordered-type double
perovskite structure \citep{aimia14} (Fig.~\ref{fig:fig1}), very
similar to centrosymmetric CaFeTi$_2$O$_6$. In this structure, the
Ca$^{2+}$ ions are ten-fold coordinated, Ti$^{4+}$ occupy tilted
octahedra, one half of the Mn$^{2+}$ are tetrahedrally coordinated
(Mn1), and the other half are in a pseudo-square planar configuration
(Mn2). However, in contrast to CaFeTi$_2$O$_6$, the square-planar
Mn$^{2+}$ and the octahedrally coordinated Ti$^{4+}$ are shifted along
the $c$-axis in CaMnTi$_2$O$_6$. This breaks the center of inversion,
lowering the symmetry from space group $P4_2/nmc$ observed in
CaFeTi$_2$O$_6$ to $P4_2mc$ and generates a spontaneous polarization.
Tetrahedrally coordinated Mn$^{2+}$ and the Ca$^{2+}$ ions play almost
no role in the ferroelectricity of CaMnTi$_2$O$_6$.  \citet{aimia14}
have shown that with increasing temperature the off-centering of the
square-planar coordinated Mn$^{2+}$ and of the Ti$^{4+}$ decreases.
Concomitantly, the second harmonic generation (SHG) signal decreases
monotonically with temperature up to the Curie temperature $T_c$ =
630 K, when it becomes zero due to the ferroelectric (S. G. $P4_2mc$)
to paraelectric (S. G. $P4_2/nmc$) phase transition.  
In paralelectric CaMnTi$_2$O$_6$ the Ti$^{4+}$-ions move to the
center of their octahedra but square-planar Mn$^{2+}$ keep their shift
along $c$ with one half of the Mn$^{2+}$ ions shifting up and the other half down 
[Fig. \ref{fig:fig1} (c)]. Thus, overall the total spontaneous
polarization disappears and a center of inversion emerges in space
group $P4_2/nmc$. 


In the present work we have studied the nature of the ferroelectric to
paraelectric phase transition of CaMnTi$_2$O$_6$ by inducing it under
high pressure. We have performed SHG and Raman spectroscopic
measurements to find the pressure-induced phase transition, understand
the effect of pressure on the spontaneous polarization, and
investigate the behavior of the Raman active modes 
Finally the high-pressure
paraelectric structure of CaMnTi$_2$O$_6$ has been solved by 
X-ray diffraction (XRD).

\section{EXPERIMENTAL DETAILS}
CaMnTi$_2$O$_6$ was synthesized using pure ilmenite-type MnTiO$_3$ and
CaTiO$_3$ sealed inside a platinum capsule at 7 GPa and
1700~$^{\circ}$C in a multi-anvil press for 30 minutes following the
work by \citet{aimia14}. The multi-anvil experiment was performed at
GFZ Potsdam with a 18/11-assembly, which was calibrated at room temperature against the phase transitions in Bi metal \citep{lloyd1971,piermarini1975}. Calibrations at high temperature are based on the following phase transitions: CaGeO$_3$: garnet-perovskite \citep{susaki1985} and SiO$_2$: coesite-stishovite \citep{akaogi1995}. Stepped graphite heaters were employed and
temperatures were measured with type C thermocouples. We obtained
orange single crystals of 10-50 $\mu$m in size embedded in powder.

High-pressure SHG, Raman spectroscopy, and single crystal XRD (SXRD)
experiments were carried out using Boehler-Almax diamond-anvil cells
(DAC) \citep{boehler06} equipped with diamonds with 350-$\mu$m culets
and tungsten gaskets, indented to a thickness of 40 $\mu$m. Holes with
a diameter of 150 $\mu$m served as sample chambers. For the Raman
spectroscopy and SXRD experiments single crystals with a thickness of
$\sim$10 $\mu$m were employed. For the SHG runs we used either a
single crystal or powder pellets. SHG signals of acentric quartz and centrosymmetric Al$_2$O$_3$ 
were used as a reference to calibrate the baseline of the SHG measurements.
In all these experiments we placed
the samples inside the gasket hole together with a ruby chip to
measure the pressure \citep{maohk78} and Ne as pressure transmitting
medium.  The SHG intensities at $\lambda_{2\omega}$ = 527 nm were
measured in transmission geometry using the setup described by
\citet{bayar09}. The Raman experiment was performed in backscattering
configuration with a Renishaw (RM-1000) spectrometer equipped with a
1800 grooves/mm grating and a spectral resolution of around 2
cm$^{-1}$. The excitation source was a HeNe laser ($\lambda$ = 633 nm)
focused down to a 10-$\mu$m spot with a 20$\times$ long working
distance objective. An edge filter was used to filter the laser
allowing measurements above 130 cm$^{-1}$. The SXRD experiment was
performed at the Extreme Conditions beamline at PETRA III with a
wavelength of $\lambda$ = 0.29036 Å focused down to 3$\times$8
$\mu$m$^2$ (FWHM) with CRL mirrors and using a PerkinElmer detector
placed at 415.5 mm from the sample. The diffraction images were
collected by 0.5$^{\circ}$ $\omega$-scanning. The image format was
converted according to the procedure described by \citet{rothk13} for
further processing with the CrysAlis$^{Pro}$ software \citep{crysal}
for indexing reflections and intensity data reduction. Crystal
structures at 0.8, 2.0, 5.6, 7.0, 9.0, 11.5, 12.7, and 14.3 GPa were
solved with the Patterson method implemented in SHELXS97-2 and refined
with SHELXL97-2 \citep{shel08}.

For the synchrotron powder XRD (PXRD) experiments we employed
gas-membrane-driven DACs equipped with 300-$\mu$m culet diamonds and
Inconel gaskets indented to 35 $\mu$m in thickness with holes of 100
$\mu$m. Powder pellets were loaded together with copper as a pressure
gauge \citep{wang09} and a mixture of 4:1 methanol-ethanol was the
pressure transmitting medium. These experiments were carried out at
the MSPD beamline at the ALBA-CELLS synchrotron ($\lambda$ = 0.4246 \AA)
up to 5 GPa. The beam was focused down to 20$\times$20 $\mu$m$^2$
(FWHM) with Kirkpatrick-Baez mirrors and a Rayonix CCD detector was
placed at 300 mm from the sample. FIT2D \citep{hamme96} was used to
integrate the diffraction patterns and GSAS \citep{larso00, tobyb01}
was employed to carry out the Le Bail refinements \citep{baila05}.
For the Le Bail refinements, a Pseudo-Voigt function was employed according to 
 \citet{thompson87} in conjunction with an 
 asymmetry correction given by \citet{fing94}.

\section{RESULTS}
\subsection{Second harmonic generation}

\begin{figure}
\centering
\includegraphics[width=0.38\textwidth]{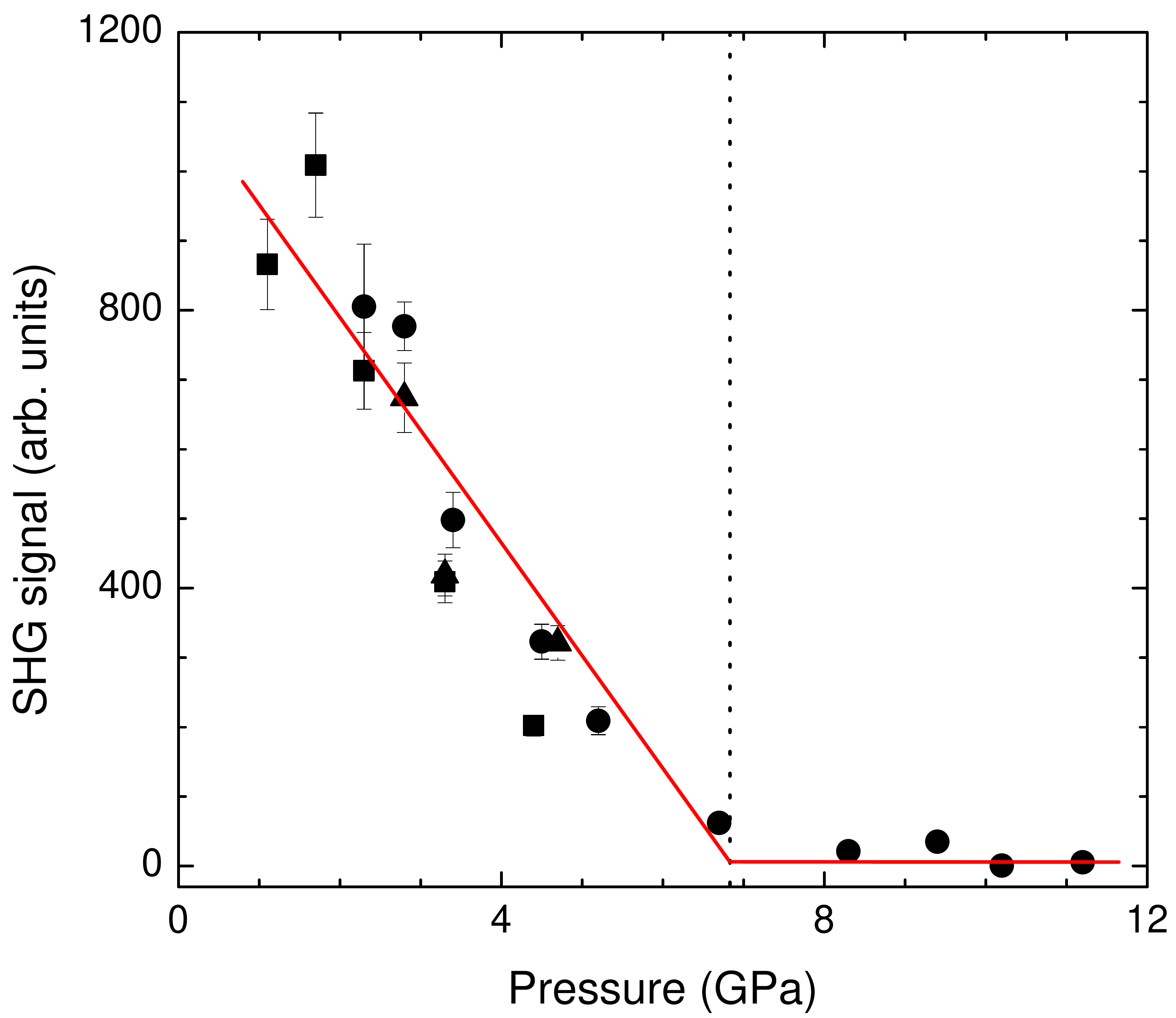}
\caption{\label{fig:fig2} Pressure dependence of the second harmonic
  generation signal showing a continuous decrease up to $\sim$7
  GPa. Different symbols correspond to different experiments. The filled
  circles are data from the single crystal experiment whereas the filled
  squares and triangles correspond to powder data. The vertical dashed
  line marks the phase boundary and the red continuous lines are
  guides for the eye.}
\end{figure}

The pressure dependence of the SHG signal $I_{2\omega}$ is shown in
Fig. \ref{fig:fig2}. With pressure, the SHG signal decreases linearly
from an initial value of $\sim$1000 counts up to 7 GPa when, within
our experimental accuracy, no signal can be detected anymore. We interpret this as the ferroelectric phase transition of CaMnTi$_2$O$_6$.

Hence, CaMnTi$_2$O$_6$ undergoes a ferroelectric to paraelectric phase transition either at
ambient pressure \citep{aimia14} and $T_c$ = 630 K or at ambient
temperature and $P_c$ = 7 GPa. This result associated to second-order phase transitions in ferroelectrics \citep{samara71} indicates that the Curie
temperature of CaMnTi$_2$O$_6$ decreases with pressure at d$T_c$/d$P =
-48$ K/GPa, a value similar to -52 K/GPa observed for BaTiO$_3$
\citep{samara71}. 
With the objective of investigating the behavior of the optical modes at the zone center and in order to solve the
high-pressure structure of CaMnTi$_2$O$_6$, we shall present in the
following sections the results of high-pressure Raman spectroscopy and
XRD studies.

\subsection{Raman spectroscopy}

\begin{figure}
\centering
\includegraphics[width=0.35\textwidth]{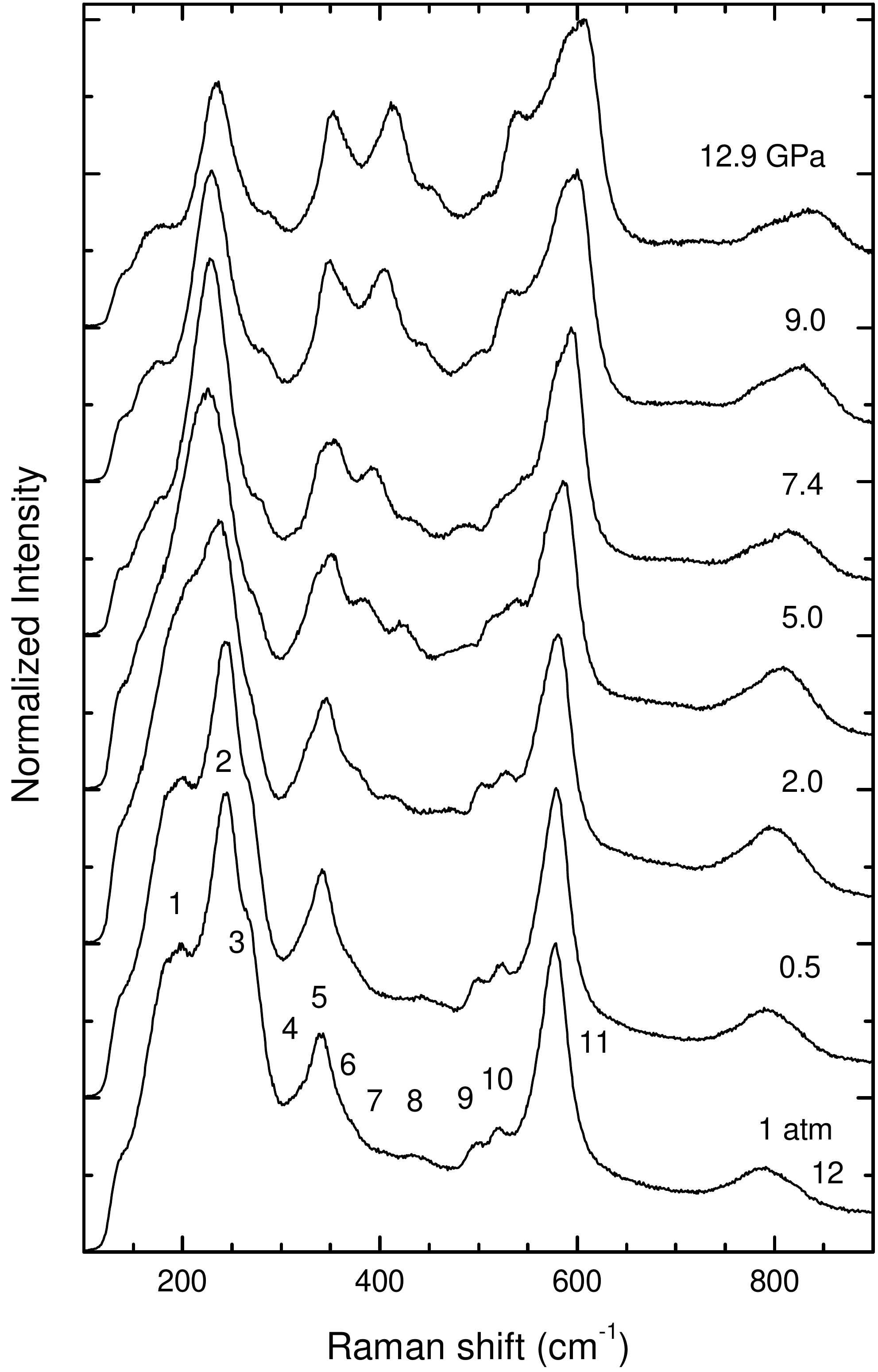}
\caption{\label{fig:fig3} Raman spectra at different pressures up to 12.9 GPa. The numbers shown in 
the spectrum at 1 atm denote the labeling of the observed Raman modes.}
\end{figure}

The A-site-ordered-type structure of CaMnTi$_2$O$_6$ with point group
$4mm$ has 88 vibrational zone-center modes. Two of them, with
irreducible representations $A_1 + E$, correspond to acoustic
phonons and the 10 $A_2$ modes are silent. The remaining optic phonons ($17 A_1 + 18 B_1 + 10 B_2 + 31 E$) are
Raman active, with the $A_1$ and $E$ modes being polar and therefore
also IR active. Given the number of Raman active modes and the fact
that 41 of them are polar and therefore will show LO-TO splitting, one
can expect a Raman spectrum with a continuous character, with a
significant overlap of Raman bands and few distinct features. A
selection of Raman spectra of CaMnTi$_2$O$_6$ at different pressures
is shown in Fig. \ref{fig:fig3}. Although an unambiguous Raman assignment
has not been possible because the size of the crystals ($<$ 50 $\mu$m)
has not allowed us to orient them, we can tentatively assign the most
intense bands.

Although CaMnTi$_2$O$_6$ has almost six times more Raman active modes
than polar ZnTiO$_3$ \citep{ruizf15} in the LiNbO$_3$-type structure, the Raman spectra of these two
compounds mutually resemble each other. This is reasonable if we consider that both, the polar crystal structures of CaMnTi$_2$O$_6$ and ZnTiO$_3$ 
are related to the perovskite structure. In case of CaMnTi$_2$O$_6$, the unit cell edge is doubled in conjunction with a distortion of the TiO$_6$ 
octahedra to form the polar, tetragonal structure in $4mm$ while a distortion of the perovskite structure according to the LiNbO$_3$ structure leads to
the polar, trigonal crystal class $3m$.

Similarly to ZnTiO$_3$, CaMnTi$_2$O$_6$ also shows two intense bands 
located at around 244 cm$^{-1}$ ($\nu_2$) and 590 cm$^{-1}$ ($\nu_{11}$), and an isolated
broad band at around 800 cm$^{-1}$ ($\nu_{12}$). In polar ZnTiO$_3$
\citep{ruizf15,caciuc00} $\nu_2$ is an $A_1$ mode polarized along the
$c$ direction and consists of an antiphase vibration of the Ti$^{4+}$
ion against the octahedral oxygen framework with the Zn$^{2+}$ at
rest, $\nu_{11}$ is another $A_1$ mode that consists of a rocking mode
of the oxygen octahedral framework with Ti$^{4+}$ and Zn$^{2+}$ at
rest, and $\nu_{12}$ is the LO mode of $\nu_{11}$. As $\nu_{2}$ and
$\nu_{11}$ are pure TiO$_6$ modes it is reasonable to assume that they
have a similar eigenvector in both compounds and therefore we
tentatively assign them to $A_1$ modes in CaMnTi$_2$O$_6$ as
well. This hypothesis is further supported under high pressure. While
eleven of the twelve observed modes shift to higher energies with
increasing pressure, the $\nu_2$ mode softens up to around 7 GPa
(Fig. \ref{fig:fig3}) when it overlaps with the mode at lower frequency $\nu_1$ and starts hardening. Above 7.4 GPa other
bands change their intensity or vanish and additional bands appear. We
interpret these changes in the Raman spectrum of CaMnTi$_2$O$_6$ at
7.4 GPa as the onset of the phase transition. This transition pressure
is in good agreement with the transition pressure deduced from the
disappearance of the SHG signal described above.

Differently to
CaMnTi$_2$O$_6$, in polar ZnTiO$_3$ \citep{ruizf15} or MnTiO$_3$
\citep{wu11} the $\nu_2$ [$A_1$ (2)] Raman mode vanishes in the
first-order phase transition while in CaMnTi$_2$O$_6$ it stays and
hardens as an indication of a less abrupt phase transformation. In fact, the behavior observed in CaMnTi$_2$O$_6$ has been previously reported in other pressure-induced second-order phase transitions \citep{errandonea2008} and indicates that even though the mode softening up to the phase transition is not driving the phase transition it is sensitive to the motion of the ions involved in the transition process. This indicates that most probably the $\nu_2$ mode softens up to the phase transition and once the Ti$^{4+}$ ion falls in the centrosymmetric site, condition to introduce the center of inversion in the structure, the same mode starts hardening with pressure. Although this hypothesis will be reinforced in the next section, such behavior can be explained as follows. Each Ti atom shares one oxygen atom with a Mn2. Under pressure the Mn2-O bond with respect to the $ab$ plane gets closer to 180$^{\circ}$ which should increase the frequency of the mode. However, the hardening is compensated below the phase transition by the softening of the mode produced by the fact that the Ti atoms are getting closer to the centrosymmetric position and therefore the vibration of the Ti atoms is facilitated. Once the Ti atoms occupy the center of inversion in the phase transition their oscillation along $c$ reduces and the hardening due to reduction of the Mn2-O angle with respect to the $ab$ plane takes over dominating the process. 

\begin{figure}
\centering
\includegraphics[width=0.35\textwidth]{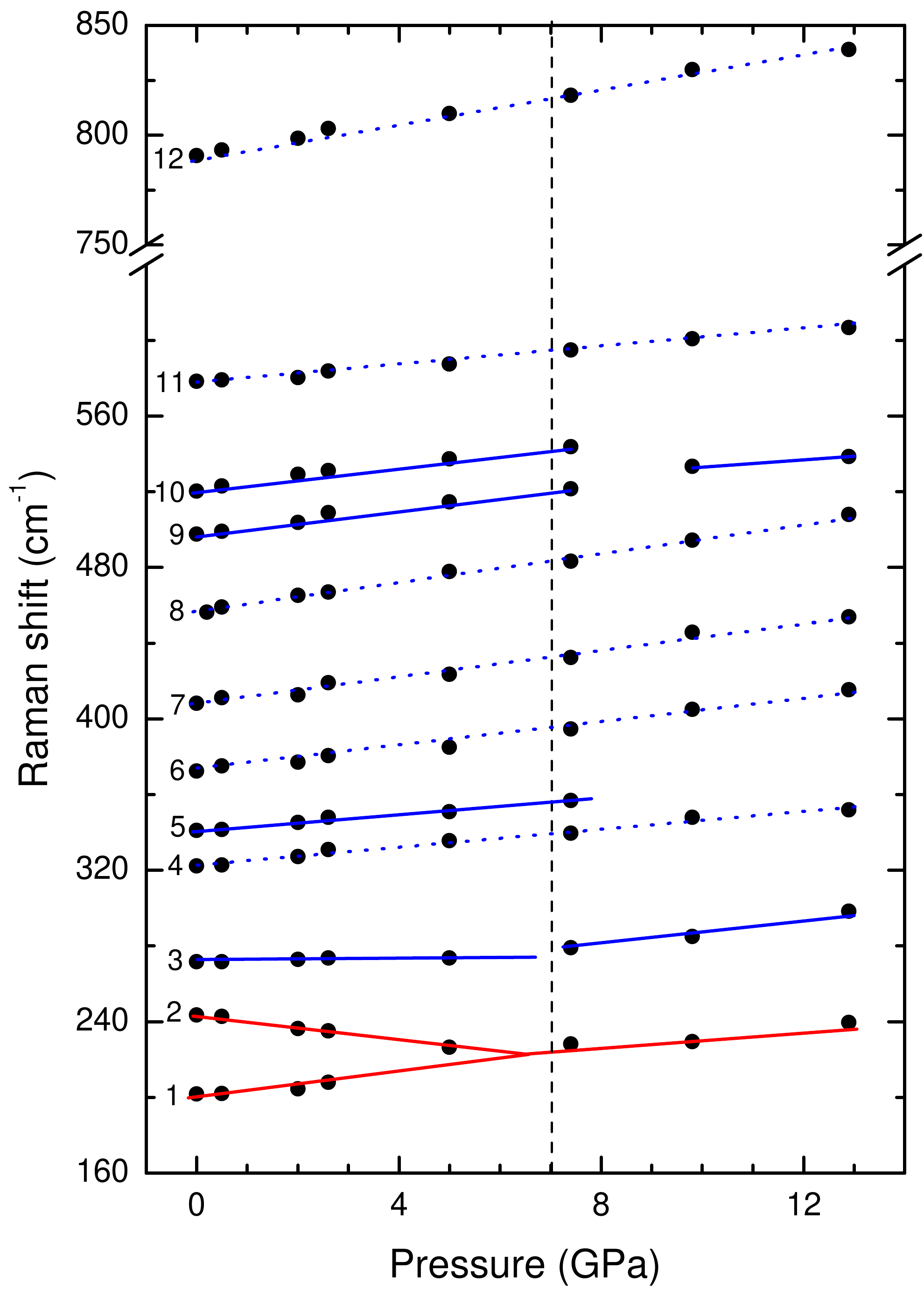}
\caption{\label{fig:fig4} Raman shifts of CaMnTi$_2$O$_6$ as a
  function of pressure measured during compression. The vertical
  dotted line marks the suggested phase transition boundary. The
  continuous and dotted blue straight lines are linear fits to the
  data. The red continuous lines mark the linear fit to the two modes
  at low frequency including the soft mode $\nu_2$. Modes are labeled
  according to Fig. \ref{fig:fig3}.}
\end{figure}

Fig. \ref{fig:fig4} shows the frequencies of the observed Raman active
modes up to 12.9 GPa. As stated before, we observe that the
frequencies of all but two modes in the low- and all modes in the
high-pressure phases increase with increasing pressure while the
frequency of the $\nu_2$ mode decreases until the onset of the phase
transition. The pressure coefficients, $a_i = \text{d}
\omega_i/\text{d} P$, obtained from linear fits are shown in Table
\ref{tab1} together with their Gr\"uneisen parameters $\gamma_i =
B_0/\omega \cdot a_i$.
 
\begin{table*}
\footnotesize
\small
\renewcommand{\arraystretch}{1.5}
\caption{Phonon frequencies of CaMnTi$_2$O$_6$ in the low-pressure
  ferroelectric and high-pressure paraelectric phases. The pressure
  coefficients $a_i$ of the modes and their Gr\"uneisen parameters $\gamma_i$
  calculated considering the bulk modulus $B_0$ = 182.95(2) GPa
  obtained from single-crystal XRD for both phases. Modes are numbered
  with increasing frequency. In the paraelectric phase those modes
  that do not have a correspondence in the ferroelectric phase are
  marked with a "$^{'}$".} \centering { \tabcolsep=0.11cm
\begin{tabular}{cccccccc}
\hline \hline     
 \multicolumn{4}{c}{\textbf{Ferroelectric at 1atm}}         & \multicolumn{4}{c}{\textbf{Paraelectric at 7.4 GPa}}                     \\  
  \hline
Mode & $\omega$    & $a_i$         & $\gamma$ & Mode & $\omega$   & $a_i$          & $\gamma$ \\
       & (cm$^{-1}$) & (cm$^{-1}$/GPa) &          &        &(cm$^{-1}$) & (cm$^{-1}$/GPa)&     \\ 
$\nu_1$  & 202              & 2.7                    & 2.45              & $\nu_2$   & 228  & 2.1  & 1.69  \\
$\nu_2$  & 244              & -3.5                    & -2.62              & $\nu_3^{'}$ & 279  & 3.6  & 2.36  \\
$\nu_3$  & 272              & 0.8                    & 0.56              & $\nu_4$       & 344  & 2.4  & 1.38       \\
$\nu_4$  & 322             & 2.4                    & 1.38             & $\nu_6$         & 395  & 3.3  & 1.62   \\
$\nu_5$  & 341             & 2.1                    & 1.13               & $\nu_7$       & 432  & 3.6  & 1.61    \\
$\nu_6$  & 372             & 3.3                   & 1.62             & $\nu_8$          & 483  & 3.9  & 1.56    \\
$\nu_7$  & 408             & 3.6                   & 1.61            & $\nu_9^{'}$       & 529  & 1.7  & 0.59   \\
$\nu_8$  & 456             & 3.9                    & 1.56             & $\nu_{11}$      & 595  & 2.3  &  0.73   \\
$\nu_9$  & 498             & 3.3                    & 1.21             & $\nu_{12}$      & 818  & 3.7  &  0.86   \\
$\nu_{10}$  & 520             & 3.1                    & 1.09             &                    \\
$\nu_{11}$  & 578             & 2.3                    & 0.73                &                        \\
$\nu_{12}$  & 791             & 3.7                    & 0.86             &                   \\
\hline \hline
\end{tabular}
}
\label{tab1}
\end{table*}

\subsection{X-ray diffraction}

\begin{figure}
\centering
\includegraphics[width=0.45\textwidth]{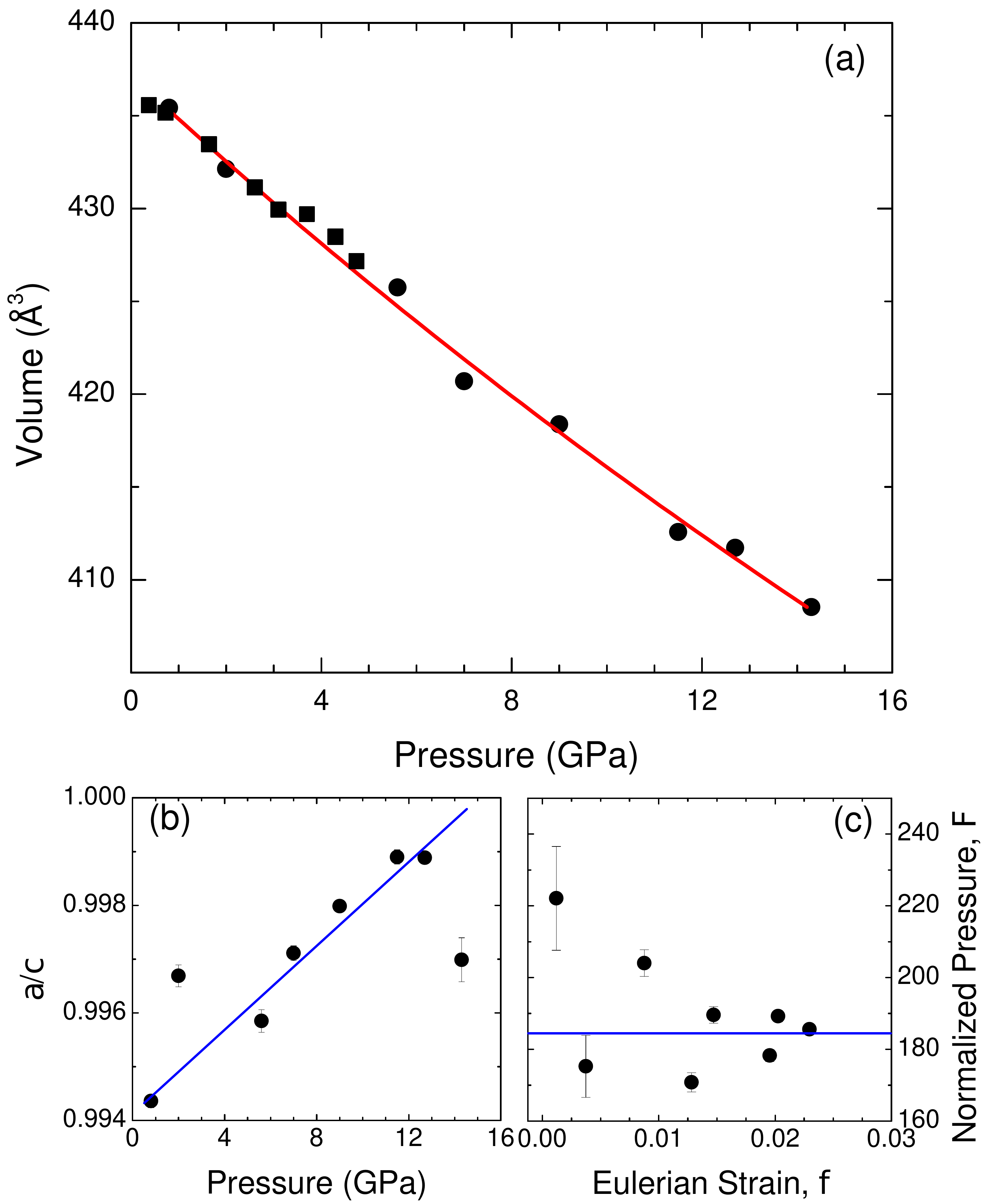}
\caption{\label{fig:fig5} (a) Pressure dependence of the unit-cell
  volume of CaMnTi$_2$O$_6$. The continuous line denotes the fit to a
  second order Birch-Murnaghan equation of state. (b) Pressure
  dependence of the $a$/$c$ ratio showing a monotonic increase. The
  continuous line is a guide for the eye. (c) Relation between the eulerian
  strain $f$ of the normalized pressure $F$. Data obtained from the
  powder diffraction experiment are represented by squares and the
  single crystal data are shown by circles.}
\end{figure}

The pressure dependence of the unit-cell volume and of the $a/c$
ratio, as well as the dependence of the normalized pressure $F$ with
the eulerian strain $f$ of CaMnTi$_2$O$_6$ are shown in
Fig. \ref{fig:fig5} for the PXRD and the SXRD experiments up to 5 and
14.3 GPa, respectively. We do not observe, within the resolution of
our experiments, any indication of a phase transition up to the
maximum pressure reached (14.3 GPa). The $a/c$ ratio of the lattice
parameters shows an increasing symmetrization of the metric with pressure. The
dependence of the normalized pressure $F$ with the eulerian strain $f$
often is a good indicator for a structural phase transition. Here,
while there is a significant scatter in our data, it seems that the
best fit yields d$F$/d$f$ = 0 and therefore all $P$-$V$ data up to
14.3 GPa are well described with a single second order Birch Murnaghan
equation of state. This is in contrast to the results obtained from
the SHG experiments and Raman spectroscopy discussed in previous
sections that indicate the occurrence of a phase transition at $\sim$7
GPa.


\begin{figure}
\centering
\includegraphics[width=0.35\textwidth]{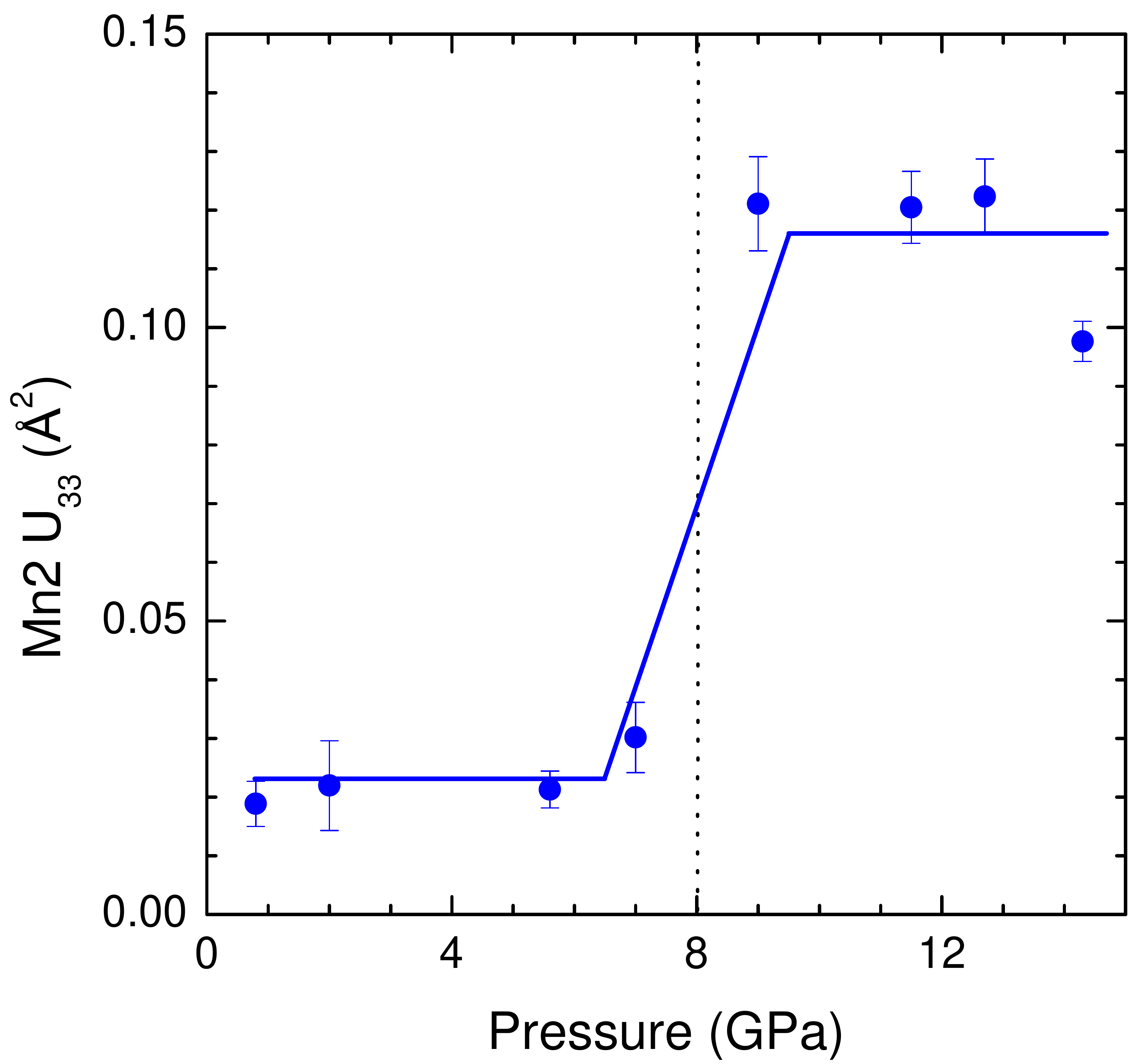}
\caption{\label{fig:fig6} Pressure dependence of the anisotropic
  $U_{33}$ thermal displacement parameter of
  Mn$^{2+}$ in the Mn2 position for the ferroelectric (S. G. $P4_2mc$)
  up to 7 GPa and the paraelectric (S. G. $P4_2mc$) structures of
  CaMnTi$_2$O$_6$ considering no splitting. Continuous lines are guides for the eye.}
\end{figure}

As we have explained in the introduction (Fig. \ref{fig:fig1}) the
structure of ferroelectric CaMnTi$_2$O$_6$ is described in
non-centrosymmetric space group $P4_2mc$, however this structure is
very similar to the structure of CaFeTi$_2$O$_6$ with the
centrosymmetric space group $P4_2/nmc$. In the eight SXRD experiments
performed at different pressures we did not find enough systematic extinctions
to distinguish between tetragonal space groups $P4_2$, $P4_2mc$, and
$P4_2/nmc$. Space group $P4_2$ is a \textsl{translationengleiche} 
subgroup of $P4_2mc$ which is simultaneously a \textsl{translationengleiche} 
subgroup of $P4_2/nmc$. In order to avoid any artifical symmetry constraints, we
initially solved the structure of CaMnTi$_2$O$_6$ at different
pressures in space group $P4_2$. We found that the structure in $P4_2$
remained identical to the structure solved by \citet{aimia14} in space
group $P4_2mc$ up to 14.3 GPa. Then we used the program PLATON
\citep{spek09} to search for higher symmetry. We found that the
symmetry of the structure was better described in $P4_2mc$ up to 7 GPa
as \citet{aimia14} had shown at ambient temperature and
pressure. However, from 9 to 14.3 GPa we found that the symmetry was
higher and the structure was better described in space group
$P4_2/nmc$ confirming the occurrence of a ferroelectric to
paraelectric phase transition as deduced from the SHG experiments
before. Using the transformation provided by PLATON we refined the
structures from 0.8 to 7 GPa in space group $P4_2mc$ and from 9 to
14.3 GPa in $P4_2/nmc$. We found that the paraelectric structure was
identical to the structure of CaFeTi$_2$O$_6$, i.e. with no off-center
shift along $c$ for Ti and Mn2. However, the $U_{33}$ term of the
anisotropic thermal displacement matrix for Mn2 increased
significantly across the phase transition. The pressure evolution of
the Mn2 $U_{33}$ is shown in Fig. \ref{fig:fig6}.

We found that up to 7 GPa $U_{33}$ takes a constant value of 0.02
Å$^2$, reasonably low considering that this is high-pressure
data. However, from 9 GPa when the structure transforms to $P4_2/nmc$
and the Mn2 atoms are constrained by symmetry to stay in the center of
the polyhedron [Fig. \ref{fig:fig1} (b)] the value of their $U_{33}$
term abruptly increases to 0.12 Å$^2$ (Fig. \ref{fig:fig6}). This
behavior is not observed for the Mn1 or for any of the other atoms and
is indicative of a large positional disorder of Mn2 along $c$. At
high temperature and ambient pressure, \citet{aimia14} have shown that
the structure of paraelectric CaMnTi$_2$O$_6$ at 700 K, though
described in space group $P4_2/nmc$, has 50\% of the Mn2 shifted up
along $c$ and the other 50\% shifted down. When we refine the
structure of CaMnTi$_2$O$_6$ with a split atom position for Mn2 with an
occupancy of 50\% we confirm that the pressure-induced ferroelectric
transformation at ambient temperature is identical to the
temperature-induced transformation at ambient pressure. \citep{supinfo}

\begin{figure}
\centering
\includegraphics[width=0.4\textwidth]{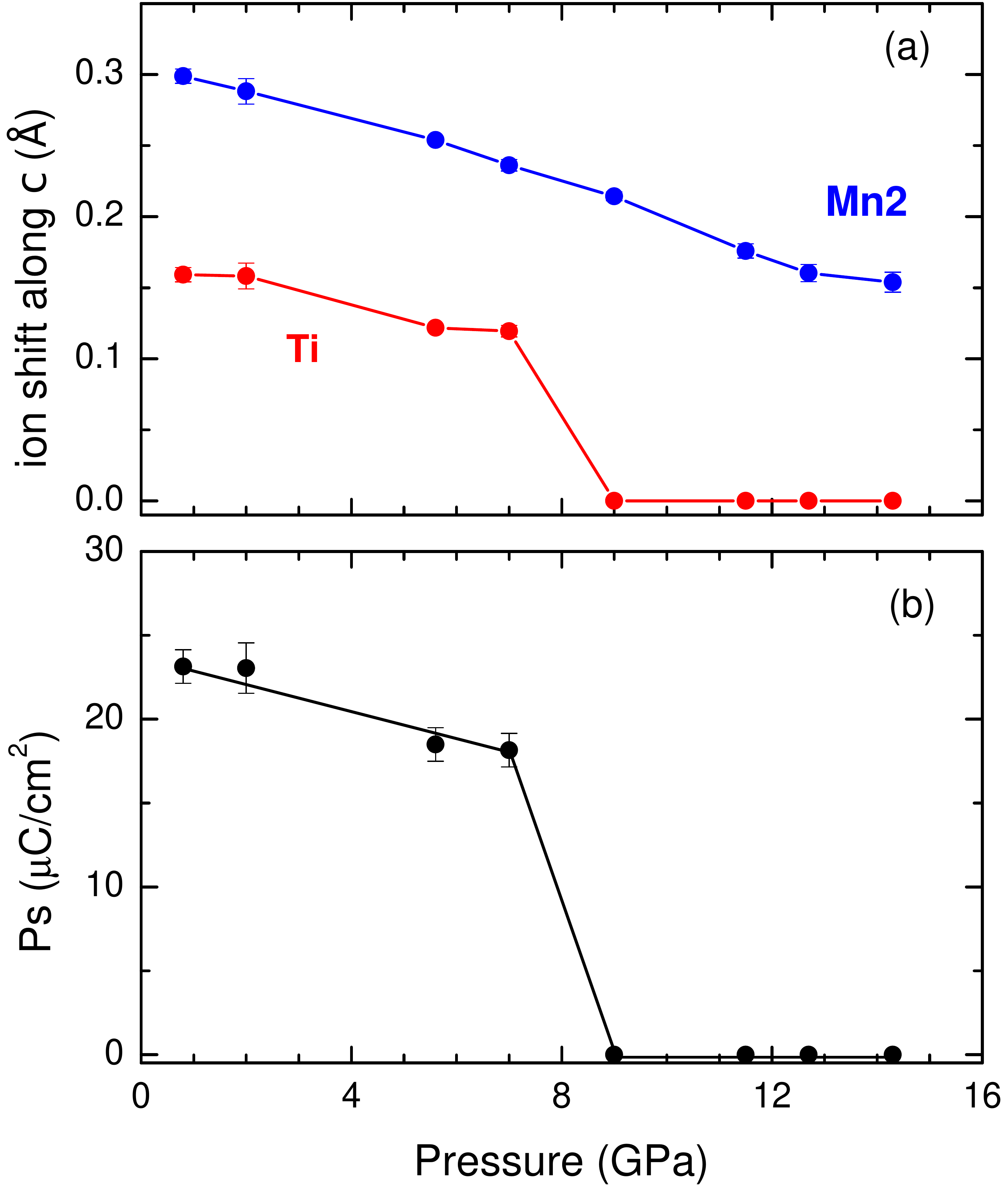}
\caption{\label{fig:fig7} (a) Off-center shifts along $c$ for Ti and
  Mn2 as a function of pressure for CaMnTi$_2$O$_6$ obtained from our
  SXRD experiments. (b) Pressure dependence of the spontaneous
  polarization $P_s$ estimated according to the point charge model
  $P_s = (\sum_iq_i\delta_i)/V$, where $q_i$ is the nominal charge of
  the atom $i$ and $\delta_i$ is the shift of the atom $i$ along $c$.}
\end{figure}

Though a temperature increase causes an expansion \citep{aimia14} of
the lattice with $\sim$2.74$\times$10$^{-5}$K$^{-1}$ and pressure
compresses the lattice, both increasing pressure or increasing
temperature leads to a reduction of the Mn2 and Ti off-center shift
along $c$ and therefore to a reduction of the spontaneous polarization
$P_s$ (Fig. \ref{fig:fig7}). Such an identical response to both pressure
or temperature increase is the result of a phase transition driven only by temperature.

Our result shows that the phase transition of CaMnTi$_2$O$_6$ is entirely driven by the off-centering of the Ti$^{4+}$
ions. There are four chemical formula units per unit cell, and hence
there are 8 Ti$^{4+}$ ions and only 2 Mn$^{2+}$ ions on the Mn2
site. Thus, when the shift of the Ti$^{4+}$ ion gets close enough to
the centrosymmetric position, i.e. a shift along $c$ of $\sim$0.1 Å
(Fig. \ref{fig:fig6}), the phase transition occurs forcing the
Mn$^{2+}$ ion in the Mn2 site (still with a shift of $\sim$ 0.2 Å) to quench
its shift resulting into a disordered Mn2 position with the Mn2
keeping the distortion. 

\section{CONCLUSIONS}
With a combined pressure-dependent study of the second harmonic
generation signal intensity, Raman spectroscopy, and single crystal
and powder x-ray diffraction we have shown that CaMnTi$_2$O$_6$
undergoes a phase transition at $\sim$ 7 GPa and ambient temperature
to the same paraelectric structure observed above 630 K and ambient
pressure.  This has allowed us to determine that the Curie temperature
decreases with pressure at d$T_c$/d$P$ = $-48$ K/GPa.  We have found
with Raman spectroscopy a mode at zone center at 244 cm$^{-1}$ that
softens with pressure. We have tentatively assigned this mode to an
antiphase vibration of the Ti$^{4+}$ ion against its octahedral oxygen
framework, indicating that the mode, despite not driving the phase transition, is coupled to the shift of
Ti$^{4+}$. With SXRD we have solved the structure of the high-pressure
phase of CaMnTi$_2$O$_6$ and followed the evolution of both the Ti and
Mn2 shifts with pressure. This has allowed us to observe that the
phase transition is mostly dominated by the shift of the Ti atoms with
the Mn2 atom playing a minor role. 

\section*{Acknowledgments}
J.R.-F. thanks the Juan de la Cierva Program (IJCI-2014-20513) of the Spanish MINECO, and D. Errandonea and A. Segura for fruitful discussions. N.S. acknowledges the DFG, Germany (Project RA2585/1-1). M.K.-M. thanks A. Ebert and R. Schulz for the technical support during the Multi-Anvil experiments. B.W. acknowledges the BMBF, Germany (Projects 05K10RFA and 05K13RF1). This paper was partially supported by the Spanish Ministerio de Econom\'ia y Competitividad (MINECO) under grants MAT2013-46649-C04-01/03-P, MAT2016-75586-C4-1/3-P, and No.MAT2015-71070-REDC (MALTA Consolider). Authors thank synchrotron ALBA-CELLS (Project 2016021588) for beamtime allocation at MSPD line. Parts of this research were carried out at the light source PETRA III at DESY (Project I-20160082 EC), a member of the Helmholtz Association (HGF). 


\end{document}